\documentclass{article}
\usepackage{graphics}

\begin{document}

\title{Superconducting Coherent States for an Extended Hubbard Model}
\author{ F. Madouri$^1$\thanks{Fethi.Madouri@ipeit.rnu.tn}, \hspace{0.2cm} Y. Hassouni\thanks{y-hassou@fsr.ac.ma}  and \hspace{0.2cm}  M. El Baz\thanks{moreagl@yahoo.co.uk}}
\date{}
\maketitle \center{\it Facult\'e des sciences, D\'epartement de
Physique, LPT, \linebreak Av. Ibn Battouta, B.P. 1014, Agdal,
Rabat, Morocco \linebreak and \linebreak $^1$ IPEIT, 1008
Monfleury, Tunis, Tunisia}
\begin{flushright}
{published in { \it Int. Jour. Mod. Phys. B17 (2003) 4859}}
\end{flushright}
\abstract{An extended Hubbard model with phonons is considered.
q-coherent states relative to the superconducting quantum symmetry
of the model are constructed and their properties studied. It is
shown that they can have energy expectation lower than eigenstates
constructed {\it via} conventional processes and that they exhibit
ODLRO.}


\vspace{0.5cm}

The Hubbard model \cite{hubbard}, despite its simplicity, is
certainly one of the important tools in condensed matter physics.
It is one of the simplest models to describe itinerant interacting
electrons on a lattice (for review see e.g. \cite{lieb}). It is
also believed that it may be used for the description of high
$T_c$-superconductivity.

In \cite{yang}, the so-called $\eta$-pairing mechanism of superconductivity was introduced. This allows the construction of states exhibiting off-diagonal long range order (ODLRO) which is regarded as a definition for superconductivity \cite{yang69}.The $\eta$ operators involved in this construction generate an $su(2)$ algebra. More precisely, it has been shown \cite{yang&zhang} that the standard Hubbard model have a $SO(4) = SU(2) \times SU(2) /Z_2 $ symmetry. The first $SU(2)$ symmetry (spin symmetry) is useful for the description of the antiferromagnetic properties of the electron systems. The second $SU(2)$ symmetry (pseudo spin), when broken, is responsible of the superconductivity of the model.

In \cite{montorsi&rasetti}, an extended Hubbard model with phonons
was shown to violate the superconducting $SU(2)$ symmetry and that
it can be only restored as a quantum (deformed) $U_q\big(
su(2)\big)$ symmetry.

In this letter, motivated by an idea \cite{penson&solomon} that
coherent states \cite{klauder, perelomov,gazeau} may be very
relevant for the study of the standard Hubbard model, we introduce
q-coherent states related to the extended Hubbard model of
\cite{montorsi&rasetti}. These states are nothing but the coherent
states related to the quantum algebra $U_q\big( su(2)\big)$ (see
e.g. \cite{ellinas}) and are shown to have, under some conditions,
energy expectation lower than states constructed through the
$\eta$-pairing mechanism and even from the Hamiltonian eigenstates
constructed in \cite{montorsi&rasetti}. We also show that this
q-coherent states exhibit ODLRO and are thus superconducting.

First, we begin by some preliminaries about the Hubbard model and recall the extended one.

\vspace{0.5cm}

Consider a D-dimensional lattice, $\Lambda$, with $L$ sites. The standard Hubbard Hamiltonian \cite{hubbard} describing a system of itinerant interacting electrons in $\Lambda$, written in a grand-canonical formalism, is
\begin{equation}
H' = H_{el}^{(loc)} + H_{el}^{(non-loc)} \label{standardH}
\end{equation}
where
\begin{eqnarray}
H_{el}^{(loc)} &=& \sum_{i\in \Lambda} \Big( -\mu (n_{i,\uparrow} + n_{i,\downarrow}) + u \; n_{i,\uparrow}n_{i,\downarrow} \Big) , \label{Hel} \\
H_{el}^{(non-loc)} &=& {1 \over 2} t \; \sum_{<i,j>} \; \sum_{\sigma = \uparrow , \downarrow} \Big( a^+_{i,\sigma} a_{j,\sigma} + a^+_{j,\sigma} a_{i, \sigma} \Big) .
\end{eqnarray}
We have used the conventional notations, namely,
\begin{itemize}
\item $ a_{i, \sigma}$ and $a^+_{i,\sigma}$ are the canonical Fermi operators describing electrons on the lattice $\Lambda$ and obey the usual anticommutation relations.
\item $i$ and $j$ are (real space) sites in $\Lambda$ and $<i,j>$ means that the summation is carried over nearest neighbor sites.
\item $\sigma = \downarrow , \uparrow$ denoting spin down and spin up respectively
\item $n_{i,\sigma} = a^+_{i,\sigma}a_{i,\sigma}$ denotes the number operator of electrons with spin $\sigma$ at site $i$.
\item $\mu$ is the chemical potential, $u$ the on site repulsion energy and t the hopping amplitude.
\end{itemize}

The model described by (\ref{standardH}) has a natural $SU(2)
\times SU(2) /Z_2$ symmetry \cite{yang&zhang} at half filling. The
first $SU(2)$ symmetry, usually called {\it spin} or {\it magnetic
symmetry}, is generated by the following {\it spin} operators
\begin{equation}
S^+ = \sum_{i\in \Lambda} a^+_{i,\uparrow} a_{i,\downarrow}\; ,\;\;\; S^- = (S^+)^{\dag} \; , \;\;\; S^z = {1 \over 2} \sum_{i\in \Lambda} (n_{j,\uparrow} - n_{j,\downarrow}) \label{magnetic}
\end{equation}
and is reminiscent of the antiferromagnetic properties of the system, see e.g. \cite{lieb}.

The second $SU(2)$ symmetry is generated by the $\eta -pairing$ operators \cite{yang}
\begin{equation}
\eta^+ = \sum_{i\in \Lambda} a^+_{i,\downarrow} a^+_{i,\uparrow} \; , \;\;\; \eta = (\eta^+)^{\dag}\; , \;\;\; \eta^z =  {1\over 2} (N-L), \label{superconducting}
\end{equation}
where $ \displaystyle N= \sum_{i\in \Lambda} n_i := \sum_{i\in \Lambda} \sum_{\sigma = \uparrow , \downarrow} n_{i,\sigma} = \sum_{i\in \Lambda} n_{i,\uparrow} + n_{i, \downarrow}$ denotes the total number operator.

It is this latter symmetry that, when broken, leads to superconducting $\eta$-paired eigenstates of the Hamiltonian (for this reason it is called a superconducting symmetry).

Both symmetries can still be conserved  when some additional terms and interactions are added to the Hamiltonian (\ref{standardH}), see e.g. \cite{A,B,C,D,E}. However this is not the case in \cite{montorsi&rasetti}. It is indeed, the latter (superconducting) $SU(2)$ symmetry that is modified (deformed) there. Their reasoning is the following.

\vspace{0.5cm}

Montorsi and Rasetti considered the extension of the Hubbard model, described by (\ref{standardH}), by including phonons in it to describe the ion vibrations. The relevant Hubbard Hamiltonian then is
\begin{equation}
H = H_{el}^{(loc)} + H_{ph}^{(loc)} + H_{el-ph}^{(loc)} + H^{(non-loc)}. \label{H}
\end{equation}
$H_{el}^{(loc)}$ is given in (\ref{Hel}), the other terms are
\begin{eqnarray}
H_{ph}^{(loc)} &=& \sum_{i\in \Lambda} \; {p_i^2\over 2M} + {1\over 2} M \omega^2 x_i^2 \; , \label{Hph}\\
H_{el-ph}^{(loc)} &=& - \lambda \sum_{i\in \Lambda} \; (n_{i,\uparrow} + n_{i,\downarrow})x_i \; , \label{Hel-ph} \\
H^{(non-loc)} &=& \sum_{<i,j>} \sum_{\sigma = \uparrow , \downarrow} \; (t_{ij} a^+_{j,\sigma}a_{i,\sigma} + h.c.),
\end{eqnarray}
where the hopping amplitude depends now on the ion displacement and is given by
\begin{equation}
t_{ij} = t \; \exp{\Big\{ \zeta (x_i - x_j)\Big\}} \; \exp{ \Big\{ ik(p_i - p_j)\Big\}}.
\end{equation}
$p_i$ and $x_i$ are local momentum and displacement operators for an ion on site $i$, $[x_i , p_j] = i\hbar \delta_{ij}$, and are both assumed to commute with the fermionic operators. (\ref{Hph}) gives the kinetic term for phonons, considered as a set of decoupled Einstein oscillators with frequency $\omega$ and mass $M$. Finally, $\lambda$ is the coupling constant for the local phonon-electron interaction (\ref{Hel-ph}).

The important fact to notice here is that when $\lambda \neq 0$
the magnetic symmetry (\ref{magnetic}) is conserved while the
superconducting (\ref{superconducting}) one is broken. It can be
restored only as a quantum (deformed) symmetry, i.e. $U_q\big(
su(2)\big) $, as proven in \cite{montorsi&rasetti} (see also
\cite{biancha&schupp}), the reasoning being basically the same as
the one in \cite{yang&zhang}. Namely, first a local $U_q\big
(su(2)\big)$ symmetry (i.e. a symmetry for the local part of the
Hamiltonian) is constructed, it is generated by the following
operators
\begin{equation}
K_i^(+) = \hbox{e}^{-i\phi p_i} a^+_{i,\uparrow} a^+_{i,\downarrow} \; , \;\;\; K_i^{(-)} = \big( K_i^{(+)}\big)^\dag \; , \;\;\; K_i^{(z)} = {1\over 2} (n_{i,\uparrow} + n_{i,\downarrow} -1 \big)\; . \label{Uqsu2}
\end{equation}
For the moment $\phi$ is an arbitrary real parameter. The operators (\ref{Uqsu2}) obey the usual $U_q\big( su(2)\big)$ commutation relations (see e.g. \cite{drinfeld, jimbo}):
\begin{equation}
[K_i^{(z)} , K_i^{(\pm )} ] = \pm K_i^{(\pm )} \; , \;\;\; [K_i^{(+)} , K_i^{(-)}] = [2K_i^{(z)}]_q \; ,
\end{equation}
with $q= \hbox{e}^{-\alpha}$, $\alpha$ being an arbitrary complex. The box function$[ ]_q$ is defined by
\begin{equation}
[X]_q = {q^X - q^{-X} \over q-q^{-1}} = {\sinh (\alpha X) \over \sinh(\alpha)} \; .
\end{equation}
For latter use we also define the q-factorial function as $[n]_q! = [n]_q [n-1]_q ...[1]_q$ , $[0]_q!=1$ for $n$ being an integer.

The requirement that the operators (\ref{Uqsu2}) commute with the local Hamiltonian $H_i^{(loc)}$ yields the following constraints on the parameters involved
\begin{equation}
\mu = {1 \over 2} \Big( u - {2 \lambda \over M \omega^2}\Big) \; , \;\;\;\; \phi = {2\lambda \over \hbar M \omega^2} \; .
\end{equation}
The first constraint being the analogue of the half-filling requirement in the standard Hubbard model (i.e. $\lambda =0$). The second one constraining the choice of the parameter $\phi$.

The extension of this local symmetry to a global one over $\Lambda$ is achieved by means of the corresponding coproduct \cite{montorsi&rasetti}. Note that, to be able to do this, it is necessary to adopt some ordering of lattice sites.

The resultant (global) $U_q\big(su(2)\big)$ algebra is generated by
\begin{eqnarray}
K^(z) &=& \sum_{i\in \Lambda} K_i ^{(z)} \label{Kz} \\
K^{(+)} &=& \sum_{j\in \Lambda} \hbox{e}^{i {\bf G.j}} \;\,  \prod_{k<j}\hbox{e}^{\alpha K_k^{(z)}} \, K_j^{(+)}  \, \prod_{k>j}\hbox{e}^{-\alpha^* K_k^{(z)}} \label{K+} \\
K^{(-)} &=& \big( K^{(+)} \big)^\dag. \label{K-}
\end{eqnarray}
The phase factor $\hbox{e}^{i {\bf G.j}}$, with $\bf{G} = (\pi ,..., \pi)$, will be useful in what follows.

The above quantum algebra $U_q\big(su(2)\big)$ describes a global symmetry of the model (i.e. the generators (\ref{Kz}-\ref{K-}) commute with the total Hamiltonian (\ref{H}) if and only if
\begin{equation}
Re(\alpha ) = {2\zeta k \over \hbar } \;\;\;\hbox{and} \;\;\;\;\; k=2\phi \; , \label{constraints2}
\end{equation}
in addition to these two constraints, in \cite{biancha&schupp} it has been shown that there are also additional constraints on the ordering of the lattice sites for the symmetry to hold. The authors of \cite{biancha&schupp} thus concluded that this symmetry holds only on one-dimensional lattices.

The first relation in (\ref{constraints2}) gives a constraint on
the real part of $\alpha$, so that its complex part can be chosen
arbitrarily, without loss of generality  we will take it null. The
second relation, mainly, means that we can interpret $\phi\over 2$
as the parameter of a Lang-Firsov transformation
\cite{biancha&schupp}.

\vspace{0.5cm}

In their paper \cite{montorsi&rasetti} Montorsi and Rasetti also
constructed eigenstates for the Hamiltonian (\ref{H}) with energy
expectation lower (or at least equal) to that of states
constructed {\it via} the conventional $\eta$-pairing mechanism.
These states are constructed from the vacuum state by successive
action of the operator $K^{(+)}$:
\begin{equation}
|\phi_n> \; = \; \beta(n,L) \; \big(K^{(+)}\big)^n \, |vac> \; , \label{phin}
\end{equation}
where the normalization constant is given by
\begin{equation}
\beta(n,L) \; = \; \Big( {[L-n]_q! \over [L]_q! \, [n]_q! } \Big)^{1\over 2} \;.
\end{equation}
This is the case when the $U_q\big(su(2)\big)$ symmetry holds.

When the model parameters have values such that this symmetry does not hold any more, one can still prove the following relation
\begin{equation}
[H,K^{(\pm)}] \; = \; \pm E K^{(\pm)} \; , \label{HK+-}
\end{equation}
where
\begin{equation}
E \; = \; u - 2\mu - 4 {\lambda^2 \over M \omega^2} \; . \label{E}
\end{equation}

\vspace{0.5cm}

Now that we have reminded all the necessar materials, we can start
developing our idea. This is an adaptation of an idea anticipated
by Solomon and Penson in \cite{penson&solomon} for a standard
Hubbard model ($SU(2)$ symmetry). In fact, they have used coherent
(pairing) states in studying the Hubbard model and proved that
this states can be useful for the description of the model. In the
following, we shall follow a similar reasoning and use q-coherent
states for the description of Montorsi and Rasetti's Hubbard
model.

Since this model have $U_q\big(su(2)\big)$ symmetry it is natural to use q-coherent states, for instance see \cite{ellinas} and references therein.

The definition we adopt for this states is the following
\begin{equation}
|\nu> \; = \; {\cal N}^{-{1\over 2}}(|\nu |^2) \;\;  \hbox{exp}_q(\nu K^{(-)}) \; |\phi_L> \; . \label{qCS}
\end{equation}
Note that we have used the maximal state
\begin{equation}
|\phi_L> \; = \; {1 \over [L]_q!} (K^{(+)})^L \; |vac> \; ,
\end{equation}
and the deformed exponential function
\begin{equation}
\hbox{exp}_q \; = \; \sum_{n=0} {x^n \over [n]_q!} \; .
\end{equation}

The normalization constant in (\ref{qCS}) can easily be computed
using some standard q-series technics \cite{gasper} and is given
by
\begin{eqnarray}
{\cal N}(|\nu|^2) &=& \big( 1 + |\nu|^2 \big)_q^L  \\
&:=& \prod_{k=1}^L \big( 1 + q^{L-2k+1} |\nu|^2 \big) \nonumber \\
&=& \sum_{n=0}^L |\nu|^{2n} \Big[_{\; n}^{\; L} \, \Big]_q  \; , \nonumber
\end{eqnarray}
where the q-binomial is defined by
\begin{equation}
\Big[_{\; n}^{\; L} \, \Big]_q \; = \; { [m]_q! \over [n]_q! [m-n]_q!} \; .
\end{equation}

The states $|\nu >$ are clearly not eigenstates of the Hamiltonian  (\ref{H}) since they involve eigenstates ($|\phi_n>$) with different $n$'s. They posses, however, many interesting properties that we shall derive in the following.

\vspace{0.5cm}

First of all, let us notice that in the limit $L \longrightarrow \infty$, $|\nu >$ is an eigenstate of ${K^{(-)} \over \sqrt{[L]_q} }$. Thus, it ($|\nu >$) can be seen as a q-harmonic oscillator coherent state, see e.g. \cite{ROMP}. This is not surprising since in the standard Hubbard model the coherent states constructed  in \cite{penson&solomon} behave similarly, i.e. in that limit their coherent state is seen as a (standard) harmonic oscillator coherent state.

\vspace{0.5cm}

Let us evaluate the energy expectation in a q-coherent state (\ref{qCS}). For this purpose we will use the following formulae, derived from (\ref{HK+-} - \ref{E}),
\begin{equation}
\begin{array}{rcl}
[H,(K^{(\pm )})^n] &=& \pm n \, E \, (K^{(\pm )} )^n \; ,\cr
[H, \hbox{exp}_q(\nu K^{(-)})] &=& - \nu \, E \, K^{(-)} \hbox{exp}_q(\nu K^{(-)}) \; . \cr
\end{array}
\end{equation}
We obtain the following formula for the desired expectation value
\begin{equation}
<\nu |H|\nu > = {\cal N} ^{-1}(|\nu |^2 ) \; E \; \sum_{n=0}^L |\nu |^{2n}  \Big[_{\; n}^{\; L} \, \Big]_q (L-n) \label{nuHnu}
\end{equation}
or, equivalently,
\begin{equation}
<\nu |H|\nu > = LE -  {\cal N} ^{-1}(|\nu |^2 ) \; E \; \sum_{n=0}^L |\nu |^{2n} n \Big[_{\; n}^{\; L} \, \Big]_q
\end{equation}

In contrast with the results of \cite{penson&solomon}, this formula for the energy expectation can not be further simplified. However, it is easy to see its behavior as the different parameters involved change. For instance, in the limit $q \longrightarrow 1$, as expected,  one recovers a similar formulae to that obtained in \cite{penson&solomon}.

Also, using the fact that $[n]_q \geq n$ is always true, it is easy to see that as $q$ tends to 1 the energy expectation (\ref{nuHnu}) tends to $E$  and after a critical value of $q$ it gets even smaller than it.

The same reasoning on $|\nu |$ implies that the energy expectation decreases as $|\nu |$ gets bigger and it gets smaller than $E$ after a critical value of $|\nu |$.

All this is confirmed by the plots (see plot {\bf a}, plot {\bf b} and plot {\bf c})

\vspace{0.5cm}

Another important feature of the q-coherent states (\ref{qCS}) is that they exhibit off-diagonal long range order (ODLRO) \cite{yang69} and thus are superconducting states. In fact, the relevant off-diagonal matrix element of the reduced density matrix $\rho_2$, by considering the states $|\phi_n>$ (\ref{phin}), is equal to $<\phi_n|a^+_{s,\uparrow} a^+_{s,\downarrow} a_{r,\downarrow} a_{r,\uparrow} |\phi_n>$. And it was found \cite{montorsi&rasetti} to be  equal to
\begin{equation}
\hbox{e}^{i{\bf G} ({\bf r} - {\bf s})} \; \hbox{e}^{\alpha (N+1 - |r-s|)} {\Big[_{\; n-1}^{\;\;\;\, L} \, \Big]_q \over \Big[_{\; n}^{\; L} \, \Big]_q} \neq 0 \; .
\end{equation}

Using this result and (\ref{qCS}), one can show that the relevant expectation value to be evaluated, $<\nu |a^+_{s,\uparrow} a^+_{s,\downarrow} a_{r,\downarrow} a_{r,\uparrow} |\nu> $, equals
\begin{equation}
\hbox{e}^{i{\bf G} ({\bf r} - {\bf s})} \; \hbox{e}^{\alpha (N+1 - |r-s|)} \Big\{ \; {1 + |\nu|^2 \over |\nu|^2} \; - \; {{\cal N} ^{-1} (|\nu|^2) \over |\nu|^2} \;\Big\}
\end{equation}
which is also different from zero (for large values of $|r-s|$). Thus the q-coherent states $|\nu >$ are superconducting.

As a matter of fact, this is not surprising . It has been argued in \cite{B,D} that superconductivity based on $\eta$-pairing is a generic rather than an exotic phenomenon. On the other hand the states $|\phi_n>$ (\ref{phin}) were shown to have pairing \cite{montorsi&rasetti}. Moreover, the coherent states of \cite{penson&solomon}, which were constructed using an $\eta$-paired state,  posses ODLRO. All these, permit us to conclude that superconductivity of coherent states, constructed from states which possess pairing, is not an exotic phenomenon.

\vspace{0.5cm}

In summary, we have constructed q-coherent states for the superconducting  $U_q\big(su(2)\big)$ symmetry of an extended Hubbard model with phonons. We have shown that they have energy expectation lower than that of the eigenstates $|\phi_n>$ constructed in \cite{montorsi&rasetti}. It was also shown that these states exhibit ODLRO, thus are superconducting. We concluded that these property (ODLRO) should be expected for any coherent states constructed using pairing states.

\newpage
\begin{picture}(0,100)
\put(0,1){\mbox{\scalebox{0.7}{\includegraphics{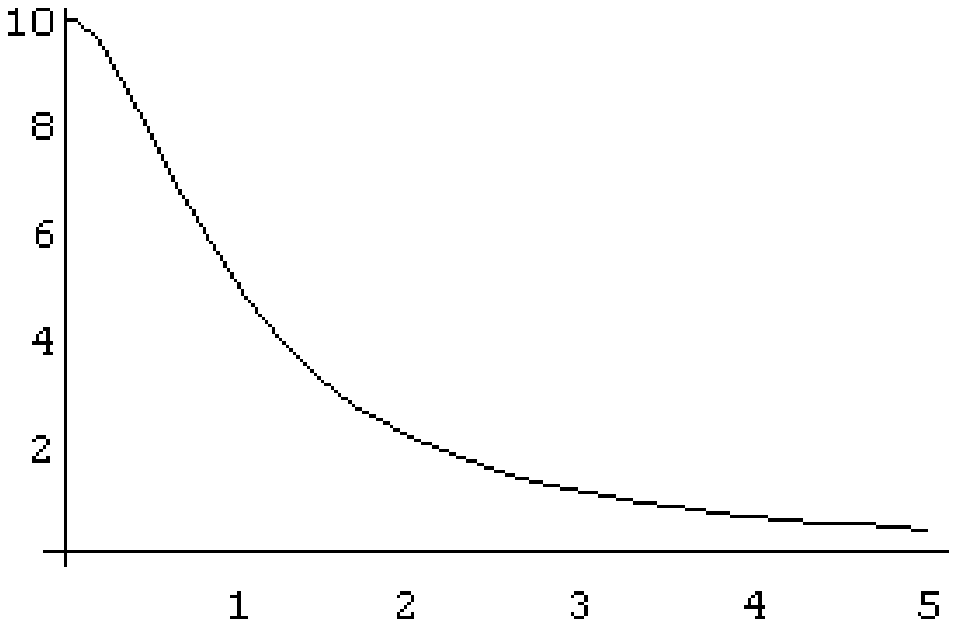}}}}
\end{picture}
{\bf Plot a.}

\begin{picture}(0,165)
\put(0,1){\mbox{\scalebox{0.7}{\includegraphics{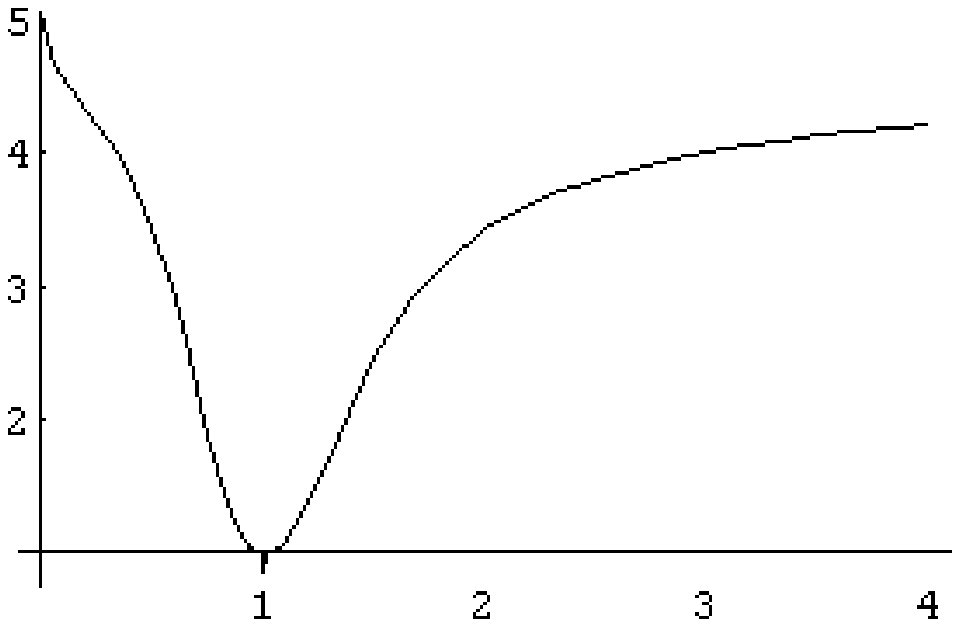}}}}
\end{picture}
{\bf Plot b.}

\begin{picture}(0,285)
\put(0,1){\mbox{\scalebox{0.7}{\includegraphics{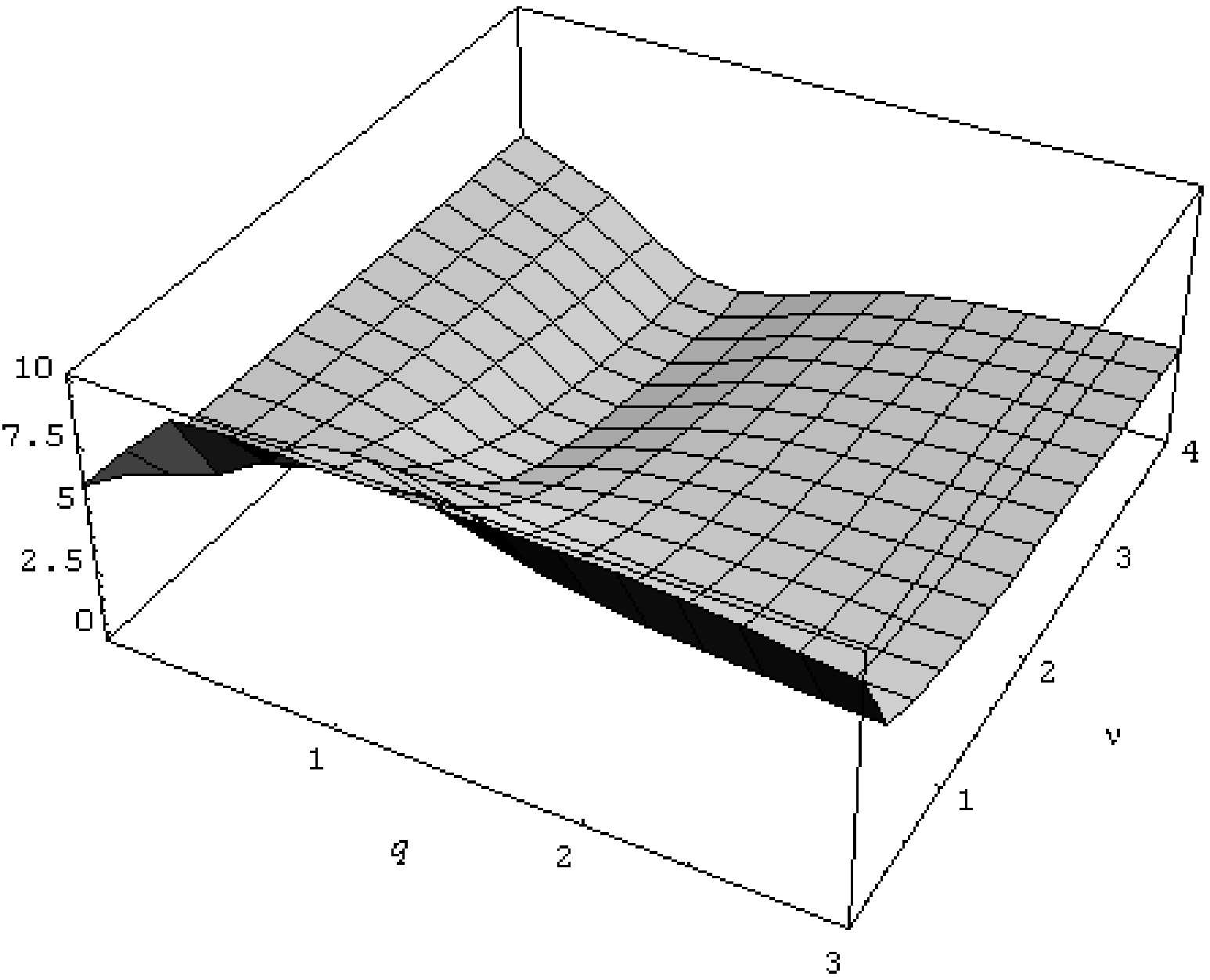}}}}
\end{picture}
{\bf Plot c.}

Energy expectation $<\nu|H|\nu>$, eq(29), versus:

- $|\nu|$ for $q=0,9$, $L=10$ and normalized $E$; {\bf Plot a.}

- $q$ for $|\nu| = 3$, $L=10$ and normalized $E$; {\bf Plot b.}

- $|\nu|$ and $q$ for $L=10$ and normalized $E$; {\bf Plotc.}
\end{document}